\documentstyle[sprocl,epsf]{article}

\def\Journal#1#2#3#4{{#1} {\bf #2}, #3 (#4)}

\def\AA{\em Acta Astronomica}
\def\ApJ{\em Astrophys. J.}
\def\ApJL{\em Astrophys. J. Lett.}

\def\be{\begin{equation}}
\def\ee{\end{equation}}
\def\bea{\begin{eqnarray}}
\def\eea{\end{eqnarray}}
\begin{document}

\title{GRAVITATIONAL RADIATION FROM MERGERS OF \\
BLACK HOLE--NEUTRON STAR BINARIES}

\author{W.H. LEE and W. KLU\'ZNIAK}

\address{University of Wisconsin, Physics Department,
\\ 1150 University Ave., Madison, WI 53706, USA}


\maketitle\abstracts{Angular momentum loss via the emission of
gravitational waves must eventually drive compact binaries containing
black holes and/or neutron stars to coalesce. The resulting events
are primary candidate sources for detectors such as VIRGO and LIGO. We
present calculations of gravitational radiation waveforms and
luminosities for the coalescence of a black hole-neutron star binary,
performed in the quadrupole approximation using a Newtonian smooth
particle hydrodynamics code. We discuss the dependence of the
waveforms and the total emitted luminosity as well as the final
configuration of the system on the initial mass ratio and the degree
of tidal locking.  }

\section{Introduction}

According to theoretical estimates$\,$\cite{lattimer,piran} some stellar-mass
black holes are expected to be formed in binary systems with neutron stars with
which they will coalesce at a rate $\ge10^{-6}$/year per galaxy.
In addition to being candidate sources for gravitational wave
detectors, such mergers have been suspected$\,$\cite{paczynski,kl} to give
rise to the observed cosmic gamma-ray bursts.\cite{fishman} We present
a summary of the results of our Newtonian hydrodynamic simulations of
the interaction of a neutron star with a black hole in a tight binary.
The results vary with the mass ratio of the two objects and depend
markedly on whether or not the rotation of the neutron star is synchronized
with the orbital motion.

We find a significant tidal transfer of angular momentum from the accreting
matter back to the mass donor---as a result the neutron star survives
mass transfer in nearly all the cases we have investigated. This greatly
extends the duration of the coalescence, the gravitational signal will not
be limited to one chirp.

\section{Numerical Method}

The three dimensional Newtonian numerical simulations presented here
have been performed using a smooth particle hydrodynamics (SPH)
code. This technique$\,$\cite{monaghan} is fully Lagrangian and
eliminates the need for a grid to carry out calculations. A tree
algorithm was implemented to optimize the computation of long-range
interactions$\,$\cite{wl}. The neutron star is modeled as a $1.4 M_{\odot}$
polytrope of index $\Gamma$=3 and unperturbed radius $R=13.4$\ km.  We
use 16,944 particles of unequal masses.  The black hole is
represented$\,$\cite{lk}
by a point mass with an absorbing spherical boundary at the
Schwarzschild radius. The gravitational radiation waveforms were
computed in the quadrupole approximation. In the dynamical
calculations presented here, radiation backreaction was not included.

\begin{figure}

{\epsfxsize=12.0cm \epsfysize=14.0cm \epsfbox{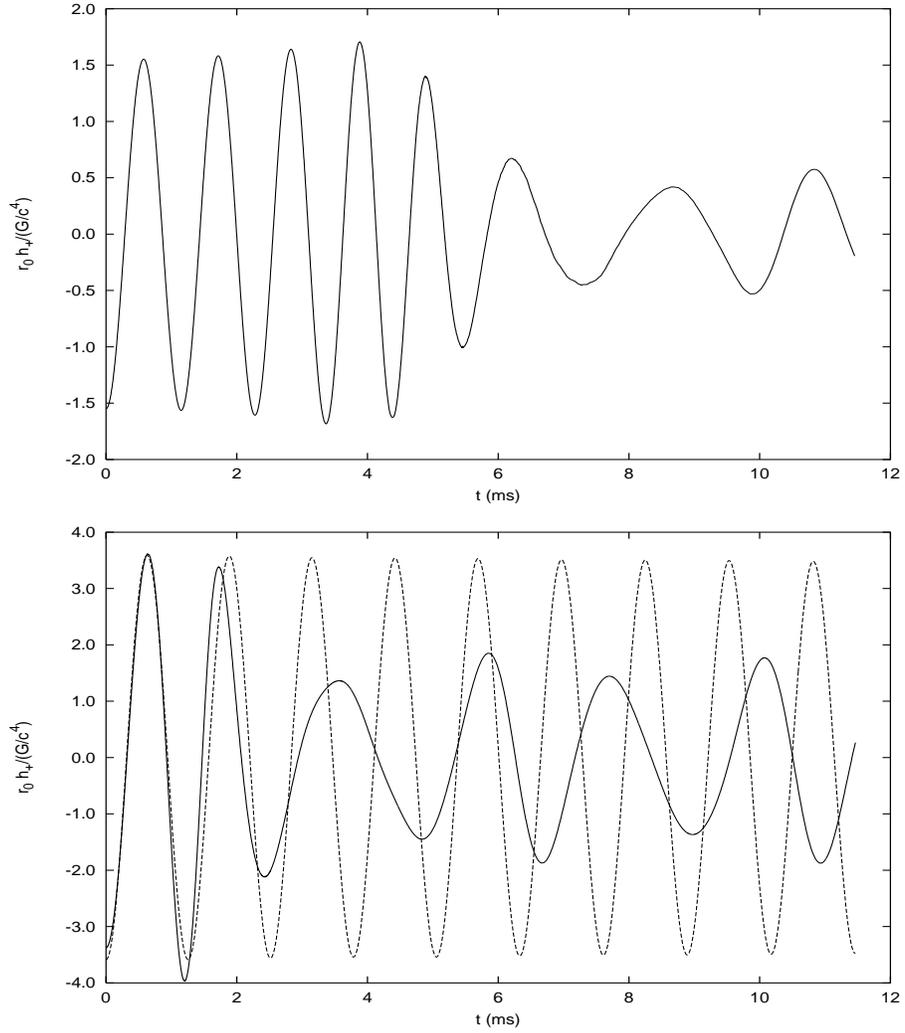}}

\caption{Gravitational radiation waveforms for an observer placed at a
distance r$_{0}$ away from the center of mass along the rotation axis of the
black hole--neutron star binary for\ \ \ 
a)~(top panel) $q=1$;
b) (bottom panel) $q=0.31$ with tidal
locking (dashed line) and no tidal locking (solid line).}

\end{figure}

\begin{figure}

{\epsfxsize=12.0cm \epsfysize=14.0cm \epsfbox{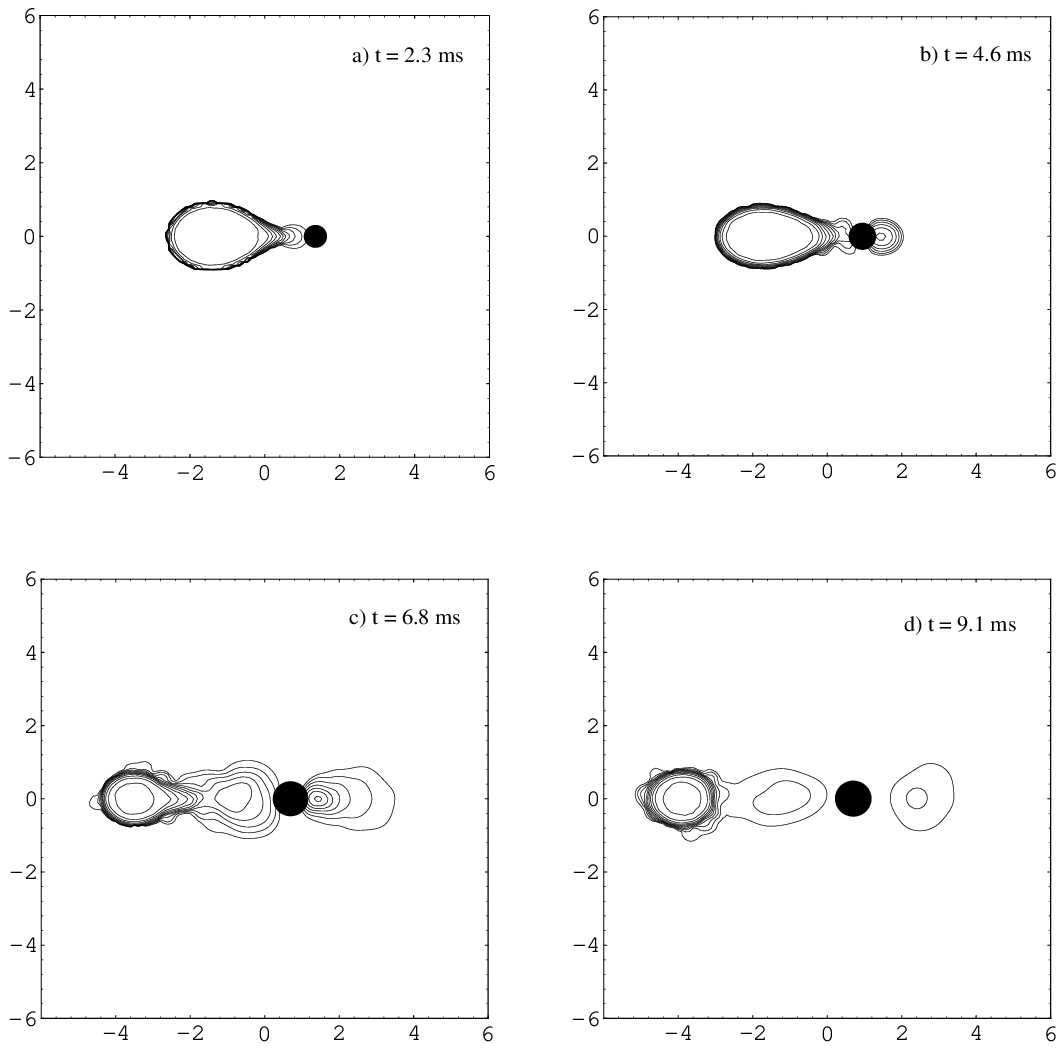}}

\caption{Density contours in the meridional plane at various times
during the dynamical coalescence of the black hole--neutron star
binary with mass ratio $q=1$. There are eleven logarithmic contours
evenly spaced every 0.25 decades, with the highest density contour at
$\rho=2.0\times 10^{14}$g cm$^{-3}$. The unit of distance is the
unperturbed stellar radius $R$. The black disk represents the black
hole (increasing in mass).}
\section*{Acknowledgments}
We gratefully acknowledge support of UNAM, KBN (grant 2P03D01311)
and the Alfred P. Sloan Foundation.

\end{figure}

\section{Results}

For a tidally locked binary with a mass ratio of unity ($q=1$)
the amplitude and frequency of the gravitational waves decrease
after the initial rise at the onset of mass
transfer (Fig. 1a). This is because  the
polytropic core survives the initial encounter with the black hole and
the final configuration of the system is an apparently stable binary
with a diminished mass ratio of $q=0.19$. 
Naturally, gravitational radiation will eventually bring the
neutron star into Roche--lobe contact again.

\begin{figure}

{\epsfxsize=12.0cm \epsfysize=14.0cm \epsfbox{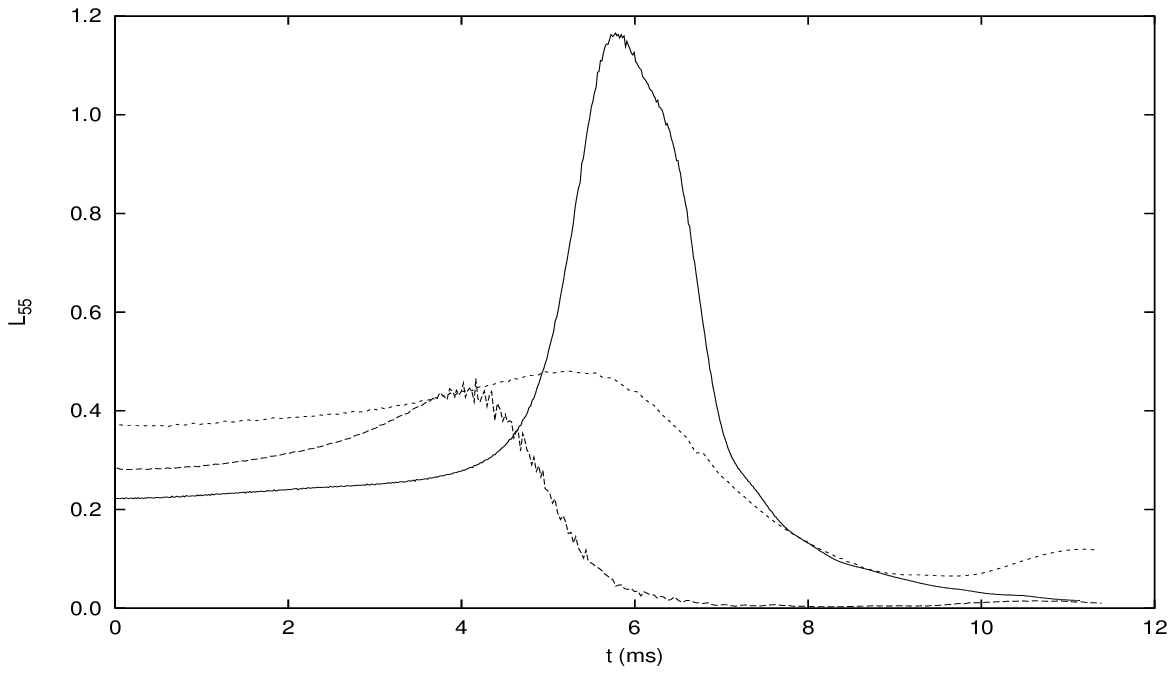}}

\caption{Gravitational wave luminosities,
$L_{55}=L/(10^{55}{\rm erg\ s}^{-1})$,
for the double neutron star
coalescence with mass ratio $q=1$ (solid line), as well as for the black
hole--neutron star coalescence with $q=1$ (dashed line) and $q=0.8$ (dotted
line).}
\end{figure}

In contrast with our preliminary results,\cite{cascina,lk} our current
simulations, performed with increased resolution,
 show that a torus lasting for a few orbits appears
around the black hole (Fig. 2). There is a
baryon--free axis present throughout the simulation (down to our
resolution of $\sim 10^{-4} M_{\odot}$).

For a lower mass ratio of $q=0.8$, again in a tidally locked binary,
the system also undergoes mass transfer on an orbital time scale, but
now no torus forms, as all the material stripped from the neutron star
is directly accreted by the black hole. Figure 3 shows the
gravitational wave luminosity for the two cases mentioned above, as
well as for a double neutron star merger.\cite{lk,rs} Note that the
three cases lead to qualitatively different results.

For an even lower mass ratio, the nature of the event is
different. There is now a brief episode of Roche--lobe overflow from
the neutron star onto the black hole, in which a modest amount of mass
is transferred and the binary separation increases.  For this tidally
locked case, one can no longer say that the hydrodynamical effects
play a predominant role, since the time scale $\tau_{GW}$ for orbital
decay due to angular momentum loss to gravitational waves is shorter
than the time of mass transfer in the present calculation.  Inclusion
of radiation reaction will change the detailed shape of the
waveforms. Nonetheless, we present (in Fig. 1b) the results for
$q=0.31$ (corresponding to a black hole mass of $4.5 M_{\odot}$) to
contrast the case of a tidally locked binary$\,$\cite{note} with the
case when the neutron star has zero spin and is initally spherical.
In the latter case, the orbital decay proceeds on a time scale
comparable to $\tau_{GW}$, there is rapid mass transfer and the
polytropic core again survives the encounter to be transferred to a
higher orbit.


\begin{thebibliography}{99}
\bibitem{note} A waveform we presented previously in fig. 3 of
ref. 2 corresponded to a $q=0.31$ configuration which had not been properly
tidally locked.
\end{thebibliography}
\end{document}